\documentclass{article}
\def\1ad{\mbox{\normalsize $^1$}}
\def\2ad{\mbox{\normalsize $^2$}}
\def\3ad{\mbox{\normalsize $^3$}}
\def\4ad{\mbox{\normalsize $^4$}}
\def\5ad{\mbox{\normalsize $^5$}}
\def\6ad{\mbox{\normalsize $^6$}}
\def\7ad{\mbox{\normalsize $^7$}}
\def\8ad{\mbox{\normalsize $^8$}}
\def\makefront{
\vspace*{1cm}\begin{center}
\def\sp{
\renewcommand{\thefootnote}{\fnsymbol{footnote}}
\footnote[1]{corresponding author~~E-mail: \email_speaker}
\renewcommand{\thefootnote}{\arabic{footnote}}
}
\def\newtitleline{\\ \vskip 5pt}
{\Large\bf\titleline}\\
\vskip 1truecm
{\large\bf\authors}\\
\vskip 5truemm
\addresses
\end{center}
\vskip 1truecm {\bf Abstract:} \abstracttext \vskip 1truecm }

\begin{document}
\def\mco{\multicolumn}
\def\epp{\epsilon^{\prime}}
\def\vep{\varepsilon}
\def\ra{\rightarrow}
\def\ppg{\pi^+\pi^-\gamma}
\def\vp{{\bf p}}
\def\ko{K^0}
\def\kb{\bar{K^0}}
\def\al{\alpha}
\def\R{{\cal R}}
\def\W{{\cal W}}
\def\L{{\cal L}}
\def\M{{\cal M}}
\def\N{{\cal N}}
\def\K{{\cal K}}
\def\E{{\cal E}}
\def\C{{\cal C}}
\def\ab{\bar{\alpha}}
\def\be{\begin{equation}}
\def\ee{\end{equation}}
\def\bea{\begin{eqnarray}}
\def\eea{\end{eqnarray}}
\newcommand{\bee}{\begin{eqnarray}}
\newcommand{\eee}{\end{eqnarray}}
\newcommand{\f}{\frac}
\newcommand{\p}{\partial}
\newcommand{\nn}{\nonumber}
\newcommand{\hy}{\hat{y}}
\newcommand{\by}{\bar{y}}
\newcommand{\bz}{\bar{z}}
\newcommand{\go}{\omega}
\newcommand{\Go}{\Omega}
\newcommand{\Gl}{\Lambda}
\newcommand{\Gpo}{{\Omega^\prime}}
\newcommand{\hsa}{$hsc^\infty (4)\,\,$}
\newcommand{\e}{\epsilon}
\newcommand{\half}{\frac{1}{2}}
\newcommand{\ga}{\alpha}
\newcommand{\gal}{\alpha}
\newcommand{\U}{\Upsilon}
\newcommand{\ups}{\upsilon}
\newcommand{\bu}{\bar{\upsilon}}
\newcommand{\dga}{{\dot{\alpha}}}
\newcommand{\dgb}{{\dot{\beta}}}
\newcommand{\gb}{\beta}
\newcommand{\gga}{\gamma}
\newcommand{\gd}{\delta}
\newcommand{\gl}{\lambda}
\newcommand{\gk}{\kappa}
\newcommand{\gep}{\epsilon}
\newcommand{\gvep}{\varepsilon}
\newcommand{\gs}{\sigma}
\newcommand{\V}{|0\rangle}
\newcommand{\ws}{\wedge\star\,}
\newcommand{\gee}{\epsilon}
\newcommand{\ggg}{\gamma}
\newcommand\ul{\underline}
\newcommand\un{{{n}}}
\newcommand\ull{{{l}}}
\newcommand\um{{{m}}}
\newcommand\ur{{{r}}}
\newcommand\us{{{s}}}
\newcommand\up{{{p}}}
\newcommand\uq{{{q}}}
\newcommand\ri{{\cal R}}
\newcommand\punc{\multiput(134,25)(15,0){5}{\line(1,0){3}}}
\newcommand\runc{\multiput(149,40)(15,0){4}{\line(1,0){3}}}
\newcommand\tunc{\multiput(164,55)(15,0){3}{\line(1,0){3}}}
\newcommand\yunc{\multiput(179,70)(15,0){2}{\line(1,0){3}}}
\newcommand\uunc{\multiput(194,85)(15,0){1}{\line(1,0){3}}}
\newcommand\aunc{\multiput(-75,15)(0,15){1}{\line(0,1){3}}}
\newcommand\sunc{\multiput(-60,15)(0,15){2}{\line(0,1){3}}}
\newcommand\dunc{\multiput(-45,15)(0,15){3}{\line(0,1){3}}}
\newcommand\func{\multiput(-30,15)(0,15){4}{\line(0,1){3}}}
\newcommand\gunc{\multiput(-15,15)(0,15){5}{\line(0,1){3}}}
\newcommand\hunc{\multiput(0,15)(0,15){6}{\line(0,1){3}}}
\newcommand\ls{\!\!\!\!\!\!\!}
\def\pa{{\parallel}}
\def\pe{{\perp}}


\def\titleline{Higher Spin Gauge Theories in Various Dimensions}

\def\email_speaker{{\tt{vasiliev@lpi.ru}}}
\def\authors{M.A.Vasiliev}
\def\addresses{I.E.Tamm Department of Theoretical Physics,
Lebedev Physical Institute,\\
Leninsky prospect 53, 119991, Moscow, Russia}

\def\abstracttext{Properties of nonlinear higher spin gauge theories
of totally symmetric massless higher spin fields in anti-de Sitter
space of any dimension are discussed with the emphasize on the
general aspects of the approach.}

\large
\makefront

\section{Introduction}
\label{Introduction}

As shown by Fronsdal \cite{Fron}, an integer-spin massless
spin-$s$ field is described by a totally symmetric tensor $\varphi
_{n _1\ldots n_s}$ ($m,n,\ldots = 0,\ldots ,d-1$ are
$d$-dimensional vector indices) subject to the double
tracelessness condition $\varphi ^{r}{}_{r}{}^{k}{}_{k n _5\ldots
n_s}=0$ which is nontrivial for $s\geq 4$. The quadratic action
for a free spin $s$ field $\varphi _{n _1\ldots n_s}$ is fixed  up
to an overall factor in the form $S_s = \varphi L \varphi $ with
some second order differential operator $L$ by the condition of
gauge invariance under the Abelian gauge transformations $
\delta\varphi_{n_1\ldots
n_s}=\partial_{\{n_1}\varepsilon_{n_2\ldots n_{s}\}} $ with
symmetric traceless tensor parameters $\varepsilon_{n _1...n
_{s-1}}$, $\varepsilon^{r}{}_{r n_3\ldots n_{s-1}}=0$. 
It is the higher spin (HS) gauge symmetry principle that makes the
HS gauge theories interesting and  perhaps  fundamental. The free
HS gauge theories extend the linearized theories of
electromagnetism (spin 1) and gravity (spin 2) in a uniform way.
The original Fronsdal theory and its more geometric versions
\cite{WF,FS} generalize the metric formulation of gravity. The HS
generalization of Cartan formulation of gravity with the HS fields
described in terms of the frame-like 1-forms was proposed in
\cite{LV,5d}. Uniformity of geometric formulations of HS fields
raises the question whether there exists an underlying nonlinear
HS gauge theory which in the free field limit gives rise to the
free Fronsdal Lagrangians. There is a number of motivations for
studying HS gauge theories.

{}From supergravity perspective, this is interesting because
theories with HS fields  may have more
supersymmetries than the ``maximal'' supergravities with 32 supercharges.
Recall that the limitation that the number of supercharges
is $\leq 32$ is a direct consequence of the requirement that
$s\leq 2$ for all fields in a supermultiplet (see e.g. \cite{PVN}).
{}From superstring perspective, the most obvious motivation is due to
Stueckelberg symmetries in the string field theory \cite{SFT},
which have a form of spontaneously broken HS gauge symmetries.
Whatever a symmetric phase of the superstring
theory is, Stueckelberg symmetries are expected to become unbroken HS
symmetries in such a phase
and the superstring field theory has to become
one or another version of the HS gauge theory.
An important indication in the same direction is that string amplitudes
exhibit certain symmetries in the high-energy limit
equivalent to the string mass parameter tending to zero  \cite{Gross}.

Unusual feature of interacting HS gauge theories is that
unbroken HS gauge symmetries do not allow flat space-time as
a vacuum solution, requiring nonzero curvature \cite{FV1}.
Anti-de Sitter ($AdS$) space is the most symmetric vacuum of this type.
This property may admit interpretation
\cite{Sund,SSO,BHS} in the context of the
$AdS/CFT$ correspondence conjecture \cite{ADSCFT}. In particular,
it was conjectured \cite{Sund,SSO} that HS gauge theories in AdS bulk
are dual to some conformal models on the
AdS boundary in the large $N$ limit with $g^2 N \to 0$
where $g^2$ is the boundary coupling constant.
Again, this indicates that  HS gauge theory has a good chance to
be related to a symmetric phase of  superstring theory.
On the other hand, a reason why the HS gauge theory may be hard to
observe in superstring theory may be that a quantum
formulation of the latter is not still available in the
$AdS$ background
despite  the progress achieved at the classical level \cite{MT}.

Whatever a motivation is, the HS problem is to find any nonlinear
theory such that
\begin{itemize}
\item
In the free field limit it contains a set of Fronsdal
fields (with correct signs of kinetic terms) plus, may be, some
other fields that admit consistent quantization
(e.g., mixed symmetry massless fields which exist in $d>4$ \cite{mixed}).
\item
HS gauge symmetries are unbroken in a nonlinear HS theory and
are deformed to some non-Abelian symmetry.
\end{itemize}
The first condition rules out ghosts\footnote{Note that
 the standard HS field dynamics is free of ghosts
but this is not the case, e.g., for the partially massless HS models
of \cite{DW} in $AdS$.}. The condition that a HS symmetry is non-Abelian
avoids a trivial possibility of building a nonlinear theory with
undeformed Abelian HS symmetries by adding
powers of gauge invariant HS field strengths to the Fronsdal action
like, for example, adding powers of the Maxwell field
strengths to a collection of free Maxwell actions instead of deforming
it to the Yang-Mills action. For the HS models we discuss,
this condition is satisfied  as a result of the manifest invariance
under diffeomorphisms. A structure of the HS symmetry is one of the
key elements of the theory.

Although being absolutely minimal, these conditions are so
restrictive that they were believed for a long period
to admit no solution at all. One argument was due to the
Coleman-Mandula type no-go theorems \cite{cm} which state that
any $S$ matrix in flat space-time, that has a symmetry larger than
(semi)direct product of usual space-time (super)symmetries and inner
symmetries, is trivial ($S= Id$). Since no scattering means no
interactions, this sounds like no theory with non-Abelian HS global
symmetries can exist.

An alternative  test was provided by the attempt \cite{diff} to
introduce interactions of a HS gauge field with gravity.
It is straightforward to see that the standard covariantization
procedure $\partial \rightarrow D=\partial -\Gamma $
breaks down the invariance under the HS gauge
transformations because, in order to prove
invariance of the action $S_s$, one has to commute derivatives,
while the commutator of the covariant derivatives is proportional
to the Riemann tensor, $[D\ldots,\, D\ldots]={\cal R}\ldots\,\, $.
As a result, the gauge variation of the
covariantized action $S_s^{cov}$ is different from zero,
having the structure
\begin{equation}
\label{fvar}
\delta S_s^{cov} =\int {\cal R}_{\ldots}(\varepsilon_{\ldots}
D\varphi_{\ldots})\neq 0\,.
\end{equation}
It seems difficult to cancel these terms because for $s >2$
they contain the Weyl tensor part of $\R$
which cannot be compensated by a
transformation of the gravitational field.

On the other hand, a number of indications on the existence of
some consistent interactions of the HS fields were found both in
the light-cone \cite{LC} and in the covariant approach \cite{cov},
giving strong evidence that some fundamental HS gauge theory must
exist. In these works, the problem was considered in flat space.
Somewhat later it was realized  \cite{FV1} that the situation
improves drastically once the problem is analyzed in the $AdS$
 space with nonzero curvature $\Lambda$. This allowed constructing
consistent $4d$ HS-gravitational interactions in the cubic
order at the action level \cite{FV1} and, later, in all orders
in interactions at the level of equations of motion \cite{d4}.
Recently, the $4d$ results of \cite{d4} were generalized to
any space-time dimension \cite{d}.

The role of $AdS$ background in HS gauge theories
is important in many respects. In particular it cancels the
Coleman-Mandula argument because  $AdS$ space  admits no
$S$-matrix, and fits naturally the AdS/CFT correspondence conjecture.
{}From the technical side, the dimensionful cosmological constant
allows new types of HS interactions with higher derivatives, which
resolve the problem with HS-gravitational interactions as follows.
The Riemann tensor ${\cal R}_{nm,kl}$ is not small near  {$AdS$}
background but
\be
\R_{nm,kl} = R_{nm,kl} -\Lambda( g_{nk}g_{ml} -g_{nl}g_{mk})\,,
\ee
where $\Lambda$ is the cosmological constant,
$g_{mn}$ is the background {$AdS$} metric tensor and $R$ is a
deviation of the Riemann tensor from the $AdS$ background curvature.
Expanding around the AdS geometry is therefore equivalent
to expanding in powers of ${ R}$ rather than in powers of the
Riemann tensor $\R$. The crucial difference compared to
the flat space is that
the commutator of covariant derivatives in the {$AdS$} space
\be
\label{DDL}
[D_n , D_m ] \sim \Lambda
\ee
is not small for general $\Lambda$.
With nonzero $\Lambda$
 one can add to the action some cubic terms schematically
written in the form
\be
\label{cact}
S^{int} =\int \sum_{p,q} \ga(p,q)\Lambda^{-\half (p+q)}
D^p (\varphi ) D^q (\varphi )
{ R}\,,
\ee
which contain higher derivatives\footnote{$D^p$ and $D^q$
denote here some combinations of derivatives of orders
$p$ and $q$, respectively.}
with some of the
coefficients proportional to negative powers of $\Lambda$.
 There exists such a unique (modulo total derivatives and field
redefinitions) action correction by the terms (\ref{cact}) with a
minimal order of derivatives that its HS gauge variation
exactly compensates the original variation (\ref{fvar})
(for any two given spins an  order of minimally necessary
derivatives in the vertex (\ref{cact}) is finite, increasing linearly
with the sum of spins). This implies that $\Lambda$ should
necessarily be nonzero in the phase with unbroken HS
gauge symmetries. In that respect, HS gauge theories are
analogous to gauged supergravities with charged gravitinos,
which also require $\Lambda \neq 0$.

Properties of  HS gauge theories are to large extent determined by
HS global symmetries of their most symmetric vacua.
HS symmetry restricts interactions and
fixes spectra of spins of massless fields in HS theories as ordinary
supersymmetry  does in supergravity.
To elucidate the structure of a global HS (super)algebra $h$ it is
useful to use the approach in which fields, action and transformation
laws are formulated in terms of the gauge fields  of $h$.
An attractive feature of this approach, which generalizes
the MacDowell-Mansouri-Stelle-West approach \cite{MM,SW} to gravity,
is that it treats all fields as differential forms with  the
gravitational field being on
 equal footing with other massless fields in a HS multiplet.
The only special property of the metric tensor is that
 it has a nonzero vacuum expectation value
allowing a meaningful linearized approximation for all fields
in the model. In section \ref{Gravity as o(d-1,2) gauge theory}
we recall the MacDowell-Mansouri-Stelle-West approach to
gravity. Then in section \ref{Higher spin gauge fields} we show
following to \cite{5d} how free HS fields can be reformulated
in terms of differential forms to be interpreted as gauge
connections. The non-Abelian HS algebra is
defined in section \ref{Higher spin  algebras} in terms of some
star-product algebra.
Then we describe unfolded formulation of free field HS dynamics in
section \ref{Unfolded higher spin dynamics} and formulate nonlinear
HS equations in section \ref{Nonlinear higher spin classical dynamics}.
Some conclusions and perspectives are summarized in the Conclusion.

\section{Gravity as o(d-1,2) gauge theory}
\label{Gravity as o(d-1,2) gauge theory}
Our approach to higher spins generalizes the
MacDowell-Mansouri-Stelle-West \cite{MM,SW} formulation of
gravity as $o(d-1,2)$ gauge theory. The key observation is
that the frame 1-form $e^a(x) = dx^\un e_\un{}^a (x)$ and
Lorentz connection $\go^{ab}(x) = dx^\un \go_\un{}^{ab} (x)$
can be interpreted as components of the  $o(d-1,2)$
connection 1-form $\go^{AB}(x) = dx^\un \go_\un{}^{AB} (x)$
($a , b, \ldots =  0,\ldots , d-1$ are fiber  Lorentz vector indices and
$A , B, \ldots =  0,\ldots , d$ are fiber $o(d-1,2)$ vector indices).
The Lorentz subalgebra $o(d-1,1)\in o(d-1,2)$ is identified  with the
stability subalgebra of some vector $V^A$. Since
Lorentz symmetry is local, this vector can be chosen
differently at different points of space-time, thus becoming a
field  $V^A = V^A (x)$. It is convenient
to relate its norm to the cosmological constant
so that $V^A$ has dimension of length\footnote
{$\Lambda$ is negative and positive in
 the {$AdS$} and dS cases, respectively (within the mostly minus
signature). For definiteness, we refer mostly
to the {$AdS$} case in this paper, although all formulae are
valid also for the dS case.}

\be
\label{vnorm}
V^AV_A = -\Lambda^{-1}\,.
\ee
This allows for a covariant definition of the
frame field and Lorentz connection \cite{SW}
\be
\label{defh}
 E^A = D(V^A) \equiv d V^A + \go^{AB}V_B\,,\qquad
\go^{\L\,AB} = \go^{AB} +\Lambda  ( E^A V^B - E^B V^A )\,. \ee
According to these definitions, $
E^A V_A \equiv 0\,, $ $ D^\L V^A = dV^A + \go^{\L\,AB}V_B \equiv
0\,. $ The theory is formulated in a way independent of a
particular choice of $V^A$. The simplest choice is with  $V^A$
being a constant vector pointing at the $(d+1)^{th} $ direction,
i.e., \be \label{Vgauge} V^A = |\Lambda|^{-1/2} \delta_d^A\,. \ee
The Lorentz directions are those orthogonal to $V^A$: $V^A A_A =0
\rightarrow A_A = A_a$. In this ``standard gauge'' the Lorentz
connection is $\go^{ab}$, and $ e^a = \go^{aB} V_B\,. $ When the
frame $E_\un^A$ has the maximal rank $d$, it gives rise to the
nondegenerate metric tensor $ g_{\un\um} = E_\un^A E_\um^B
\eta_{AB}\,. $

The $o(d-1,2)$ Yang-Mills field strength is
\be
\label{R}
R^{AB}= d\go^{AB} + \go^{A}{}_C \wedge \go^{CB}\,.
\ee
It can be decomposed into the torsion part
$
R^A\equiv DE^A\equiv  R^{AB} V_B \,
$
and the $V$--transversal Lorentz part.
In the standard gauge (\ref{Vgauge}) they identify with
the torsion tensor and Riemann tensor shifted by the terms
bilinear in the frame 1-form
\be
\label{rlon}
R^a = d e^a + \go^a{}_b \wedge e^b\,,\qquad
R^{ab}= d\go^{ab} + \go^{a}{}_c \wedge \go^{cb} +\Lambda e^a\wedge e^b \,.
\ee
The zero-torsion condition
$
R^a = 0\,
$
expresses the Lorentz connection via the frame field
in the usual manner. Provided that the metric tensor is nondegenerate,
any field $\go$ satisfying the zero-curvature equation $R^{AB}=0$
describes $(A)dS_d$ space with the cosmological constant $\Lambda$,
\be
\label{ADS}
AdS_d:\qquad R^{AB}=0\,,\quad rank|E_\un^A| = d\,.
\ee

The action of Stelle and West \cite{SW} for $4d$ gravity is
\be
\label{MM}
S=-\frac{1}{4\kappa^2 |\Lambda|^{1/2}}\int_{M^4}\varepsilon_{ABCDE}
V^A R^{BC} R^{DE}\,.
\ee
In the standard gauge  it amounts to the action of
MacDowell-Mansouri \cite{MM}. To see that it is equivalent
to the Einstein action with the cosmological term one decomposes
$R^{ab}= \R_\L^{ab} +\Lambda R_\C^{ab}$ where
$R^{ab}_\L$ is the field strength of the Lorentz subalgebra $o(d-1,1)$
and $R_\C^{ab}=e^a\wedge e^b$.
Plugging this into (\ref{MM}) one observes that the
terms $R_\L\times R_\L$ form a topological invariant (Gauss-Bonnet),
which does not contribute to the field equations, $R_\L\times R_\C$
terms form the scalar curvature
and $R_\C \times R_\C$  terms give rise to the cosmological term.

The field $V^A$ makes the $o(d-1,2)$ gauge  symmetry manifest
\be
\label{otr}
\delta \go^{AB} = D \epsilon^{AB}  \,,\qquad
\delta V^A = - \epsilon^{AB}V_B\,.
\ee
Also, the action (\ref{MM}) is manifestly invariant under
diffeomorphisms because of using the exterior algebra formalism.
Let us define a covariantized diffeomorphism transformation as a
mixture of a usual diffeomorphism, generated by an infinitesimal
vector field parameter $\xi^\un$, with a local $o(d-1,2)$ gauge
symmetry transformation with the parameter
$\gvep^{AB}_\xi =- \xi^\un\go_\un^{AB}$. The transformation law
under the covariantized diffeomorphisms is
\be
\delta_\xi V^A = \xi^\un D_\un V^A\equiv \xi^\un E_\un^A\,,
\ee
\be
\label{xir}
\delta_\xi \go_\um^{A B}=\xi^\un \partial_\un \go_\um^{A B}
+\partial_\um (\xi^\un ) \go_\un^{A B} -D_\um (\xi^\un \go_\un^{AB})\equiv
\xi^\un R_{\un\um}^{AB}\,.
\ee

Fixing some gauge for $V^A$, one relates the
parameters of diffeomorphisms and gauge transformations
from $o(d-1,2)/o(d-1,1)$ via the condition
\be
\label{vinv}
\delta V^A =0=  \xi^\un E_\un^A -  \gvep^{AB}V_B\,.
\ee
When $V^A$ is a constant vector, the condition
does not restrict the parameters of true (i.e., noncovariantized)
diffeomorphisms, rather restricting to zero the gauge parameters
from $o(d-1,2)/o(d-1,1)$.
The resulting theory turns out to be expressed in terms of
the frame field and Lorentz connection contained in
$\go^{AB}$  and, as it should, is manifestly invariant under
local Lorentz transformations and diffeomorphisms.
Alternatively, expressing the covariantized diffeomorphism
parameters from (\ref{vinv})
\be
\label{kil}
\xi^\un g_{\un\um} = E_\um{}_A \gvep^{AB}V_B
\ee
and using
(\ref{xir}), one can interpret the mixture of diffeomorphisms and
transformations from
$o(d-1,2)/o(d-1,1)$, that leave invariant $V^A$, as a deformation
of the transformation law (\ref{otr}) for the connection $\go^{AB}$
by some $R$-dependent terms.

With the help of $V_A$ it is straightforward to write a
$d$--dimensional generalization \cite{5d} of the
MacDowell-Mansouri-Stelle-West action \be \label{gact}
S=-\frac{1}{4|\Lambda|^{1/2} \kappa^{d-2}}\int_{M^d} \gep_{A_1
\ldots A_{d+1}}R^{A_1 A_2}\wedge R^{A_3 A_4} \wedge
E^{A_5}\wedge\ldots     \wedge E^{A_{d}} V^{A_{d+1}}\,. \ee
Obviously, among various solutions of the equations of motion of
the action (\ref{gact}), any flat connection with $R^{AB}=0$ is a
solution. According to (\ref{ADS}) it describes $AdS_d$ provided
that the metric is nondegenerate.  (Different flat connections
with nondegenerate metric describe $AdS_d$ in different
coordinates.) This solution is most symmetric. Indeed, let
$\go_0^{AB} (x)$ be some solution of (\ref{ADS}). According to
(\ref{xir}), covariantized diffeomorphisms act trivially on
$\go_0$. Local $o(d-1,2)$ transformations (\ref{otr}) map one
solution $\go_0$ of (\ref{ADS}) to another. The transformations,
that leave $\go_0$ invariant, satisfy \be \label{de} D_0
\gvep^{AB}(x)=0\,, \ee where $D_0$ is the $o(d-1,2)$ covariant
derivative constructed from $\go_0$. Equation (\ref{de}) is
formally consistent because $D_0^2 =0$. As a result, it determines
all derivatives of the 0-form $\gvep^{AB}(x)$ in terms of its
values $\gvep^{AB}(x_0)$ at any given point $x_0$. So, in the
topologically trivial situation, any solution $\gvep^{AB}_0 (x)$
of (\ref{de}) is fixed in terms of $\gvep^{AB}_0 (x_0) \in
o(d-1,2)$ which are arbitrary parameters of the  global symmetry
$o(d-1,2)$ of the vacuum (\ref{ADS}). It is important to note that
the same symmetry can be realized by diffeomorphisms generated by
the Killing vector fields $\xi^\un$ (\ref{kil}) of the metric
$g_{0\,nm}= E_{0\,n}^A E_{0\,m A}$ where $E_0$ is defined by
(\ref{defh}) with the $AdS$ background connection $\go_0{}^{AB}$.
This explains why the space-time symmetry algebras associated with
the motions of the most symmetric vacuum spaces reappear as gauge
symmetry algebras in the gauge approach to gravity.

The lesson is that, to describe a gauge model that has a
global symmetry $h$, it is useful to reformulate it in terms
of the gauge connections $\go$ and curvatures $R$ of $h$ in such a way that
the zero curvature equation $R=0$ solves the field equations.
If a symmetry $h$  is not known, this observation can be used other way
around  to guess what it is by reformulating
dynamics a la MacDowell-Mansouri to guess a structure of an appropriate
curvature $R$.

\section{Higher spin gauge fields}
\label{Higher spin gauge fields}

In the spin two case, the formulation in terms of gauge connections
results from the extension
$
g_{\un\um}\longrightarrow \{e_\un^a ,
\go_\un^{ab}\}\longrightarrow \go_\un^{AB}.
$
It has the following generalization to  any spin
\be
\label{arr}
s\geq 1:\qquad \varphi_{\un_1\ldots \un_s}
\longrightarrow \{e_\un{}^{a_1\ldots a_{s-1}},
\go_\un{}^{a_1\ldots a_{s-1},b_1\ldots b_t}\Big |_{t=1,2\ldots s-1}\}\,
\longrightarrow \go_\un^{A_1\ldots A_{s-1},B_1\ldots B_{s-1}}\,.
\ee

The first arrow is for the equivalent free field reformulation \cite{LV}
of the Fronsdal dynamics in terms of the set of 1-forms
$dx^\un \go_\un{}^{a_1 \ldots a_{s-1}, b_1\ldots b_t }$
$(0\leq t\leq s-1)$ which contain the frame type dynamical field
$e_\un{}^{a_1\ldots a_{s-1}}=\go_\un{}^{a_1\ldots a_{s-1}}$ ($t=0$)
and the generalized Lorentz connections
$\go^{a_1\ldots a_{s-1},b_1\ldots b_{t}}$ $(t>0)$.
The connections $\go^{a_1\ldots a_{s-1},b_1\ldots b_{t}}$ are
symmetric in the fiber Lorentz vector indices
$a_i$ and $b_j$ separately, satisfy the antisymmetry condition
\be
\label{asym}
\go_\un{}^{a_1 \ldots a_{s-1}, a_s b_2\ldots b_t }   =0 \,
\ee
implying that symmetrization over any $s$ fiber indices gives zero,
and are traceless with respect to the fiber indices
$\go_\un{}^{a_1 \ldots a_{s-3}c}{}_c{}^{, b_1\ldots b_t } =0 $,
$\go_\un{}^{a_1 \ldots a_{s-2}c}{}{}^{,}{}_c{}^{b_2\ldots b_t } =0 $,
$\go_\un{}^{a_1 \ldots a_{s-1},}{}_c{}^{c b_3\ldots b_t } =0 $.
The HS gauge fields associated with the spin $s$ massless
field therefore take values in the direct sum of all irreducible
representations of the $d$-dimensional massless Lorentz group
$o(d-1,1)$ described by the Young diagrams with at most two rows
such that the longest row has length $s-1$ \,
\begin{picture}(45,1.5)
{\linethickness{.250mm}
\put(00,00){\line(1,0){30}}%
\put(00,05){\line(1,0){40}}%
\put(00,10){\line(1,0){40}}%
\put(00,00){\line(0,1){10}}%
\put(05,00.0){\line(0,1){10}} \put(10,00.0){\line(0,1){10}}
\put(15,00.0){\line(0,1){10}} \put(20,00.0){\line(0,1){10}}
\put(25,00.0){\line(0,1){10}} \put(30,00.0){\line(0,1){10}}
\put(35,05.0){\line(0,1){05}} \put(40,05.0){\line(0,1){05}}
}
\put(12,10.9){\scriptsize  $s-1$}
\put(16,-5){\scriptsize  $t$}
\end{picture} .
Analogously to the relationship between metric
and frame formulations of the linearized gravity,
the totally symmetric double traceless HS
fields used to describe the HS dynamics in the metric type
formalism \cite{Fron,WF} identify with the symmetrized part
$\varphi_{a_1 \ldots a_s} = \go_{\{ a_1\, \ldots a_s\} }$ of the
frame type field $\go_\un{}^{a_1 \ldots a_{s-1}}$. The antisymmetric
part in $\go_\un{\,}^{a_1 \ldots a_{s-1}}$ can be gauge fixed to
zero with the aid of the generalized HS Lorentz symmetries with the
parameter $\gep{}^{a_1 \ldots a_{s-1},  b }$. That
$\varphi_{a_1 \ldots a_s}$ is double traceless is a  consequence of
the tracelessness of $\go_\un{\,}^{a_1 \ldots a_{s-1}}$ in the indices
$a_i$. The generalized Lorentz connections
$\go_\un{}^{a(s-1),b(t)}$ with $t>0$
are auxiliary fields expressed through order-$t$ derivatives of
the dynamical frame-like field by certain constrains,
$\go_\un{}^{a(s-1),b(t)}\sim \left ( \f{1}{\sqrt{\Lambda}}
\f{\p}{\p x}\right )^t (e)$.

The second arrow in (\ref{arr}) expresses the observation of
\cite{5d} that the set of the HS 1-forms $dx^\un \go_\un{}^{a_1
\ldots a_{s-1}, b_1\ldots b_t }$ with all $0\leq t \leq s-1$
results from the ``dimensional reduction'' of a 1-form $dx^\un
\go_\un{}^{A_1 \ldots A_{s-1}, B_1\ldots B_{s-1} }$ carrying the
irreducible representation of the $AdS_d$ algebra $o(d-1,2)$
described by the traceless two-row rectangular Young tableau of
length $s-1$, i.e. \be \label{lirre} \go^{\{A_1 \ldots
A_{s-1},A_s\} B_2\ldots B_{s-1} } =0\,,\qquad \go^{A_1 \ldots
A_{s-3}C}{}_{C,}{}^{B_1\ldots B_{s-1} } =0\,. \ee The Lorentz
covariant irreducible fields $dx^\un \go_\un{}^{a_1 \ldots
a_{s-1}, b_1\ldots b_t }$ identify with those components of
$dx^\un \go_\un{}^{A_1 \ldots A_{s-1}, B_1\ldots B_{s-1} }$ which
are parallel to the compensator field $V^A$ in $s-t-1$  indices
and are transversal in the rest ones.

The linearized HS curvature $R_1$ has the
form of the covariant derivative $D_0$
with the $o(d-1,2)$ connection $\go_0^{AB}$ in the representation
of $o(d-1,2)$ described by the two-row  rectangular Young tableau
of length $s-1$, i.e.
\bee
\label{R1A}
R_1^{A_1 \ldots A_{s-1}, B_1\ldots B_{s-1} } &=& D_0
(\go^{A_1 \ldots A_{s-1}, B_1\ldots B_{s-1}}) =
d \go^{A_1 \ldots A_{s-1}, B_1\ldots B_{s-1} }\nn\\
 &{}&\ls\ls\ls\ls\ls\ls\ls\ls\ls +(s-1)\Big(
\go_0^{\{A_1}{}_{C}\wedge
\go^{C A_2 \ldots A_{s-1}\}, B_1\ldots B_{s-1} }
+\go_0^{\{B_1}{}_{C}\wedge
\go^{ A_1 \ldots A_{s-1}, C B_2\ldots B_{s-1}\} }\Big ),
\eee
where $\go_0^{AB}$ is the background $AdS_d$ gauge field
satisfying the flatness condition $D_0^2=0$ (\ref{ADS})
which guarantees that
the linearized curvature (\ref{R1A}) is invariant under the
Abelian HS gauge transformations with the HS gauge parameters
$\gee^{A_1 \ldots A_{s-1}, B_1\ldots B_{s-1} }$
\bee
\label{dgo}
\delta \go_1{}^{A_1 \ldots A_{s-1}, B_1\ldots B_{s-1} }  =
D_0 \e^{A_1 \ldots A_{s-1}, B_1\ldots B_{s-1} }\,.
\eee

The $o(d-1,2)$
covariant form of the free action for a massless spin $s$ field is
\cite{5d}
\bee
\label{gcovdact}
S^s_2&=&\half
\sum_{p=0}^{s-2}a (s,p)
\gep_{A_1 \ldots A_{d+1}}\int_{M^d} E^{A_5}\wedge\ldots
\wedge E^{A_{d}}V^{A_{d+1}} V_{C_1}\ldots V_{C_{2(s-2-p)}}\nn\\
&{}&\ls\ls\ls \wedge
R_1^{A_1 B_1 \ldots B_{s-2}}{}_,{}^{A_2 C_1 \ldots C_{s-2-p}
D_1\ldots D_p}\wedge R_1^{A_3}{}_{B_1 \ldots B_{s-2},}{}^{ A_4 C_{s-1-p} \ldots
C_{2(s-2-p)}}{}_{D_1\ldots  D_p }\,,
\eee
where
\be
\label{al}
a (s,p) = \tilde {a} (s) (-\Lambda)^{-(s-p-1)}
\frac{(d-5 +2 (s-p-2))!!\, (s-p-1)}{  \,(s-p-2)!}\,.
\ee
The coefficients $a(s,p)$ are fixed up to an overall
spin-dependent factor $\tilde{a}(s)$
by the ``extra field decoupling
condition'' that the
variation of the free action (\ref{gcovdact}) is different from
zero only for the fields
$
\go_\un{}_{A_1 \ldots A_{s-1},B_1 }=
\go_\un{}_{A_1 \ldots A_{s-1} }{}_{,B_1\ldots B_{s-1}} V^{B_2} \ldots
V^{B_{s-1}}
$
which contain the frame type dynamical HS field
$\go_\un{}^{a_1 \ldots a_{s-1} }$  and the Lorentz type
auxiliary field $\go_\un{}^{a_1 \ldots a_{s-1}, b }$, which is expressed
in terms of the frame type field by virtue of
its equation of motion equivalent to the ``zero torsion condition"
$
R_{1\,A_1 \ldots A_{s-1} }{}_{,B_1\ldots B_{s-1}} V^{B_1} \ldots
V^{B_{s-1}}=0\,.
$
Insertion of the expression for
$\go_\un{}^{a_1 \ldots a_{s-1}, b }$ into (\ref{gcovdact})
gives rise to the HS action expressed entirely
(modulo total derivatives) in terms
of $\go_\un{}^{a_1 \ldots a_{s-1} }$ and its first derivatives.
Since the linearized curvature (\ref{R1A})
is  invariant under the Abelian HS gauge transformations (\ref{dgo})
the resulting action has necessary
HS gauge symmetries and, because of the extra field decoupling
condition, describes
correctly the free field HS dynamics in $AdS_d$.
In particular, the generalized Lorentz-like transformations
with the gauge parameter $\gep_{A_1 \ldots A_{s-1} }{}_{,B_1} (x)$
guarantee that only the totally symmetric Fronsdal part
$\varphi_{a_1 \ldots a_s} = \go_{\{ a_1\, \ldots a_s\} }$
of the frame type gauge field  contributes to the action.

\section{Higher spin  algebras}
\label{Higher spin  algebras}

Once dynamics of totally symmetric
HS gauge fields is shown to be described
by 1-forms taking values in the two-row rectangular Young tableaux
of $o(d-1,2)$, this suggests that a
 $AdS_d$ HS algebra $h$ admits a basis formed by a set of elements
$T_{A_1\ldots A_n ,B_1\ldots B_n}$, which satisfy the properties
analogous to (\ref{lirre}),
$T_{\{A_1 \ldots A_{s-1},A_s\} B_2\ldots B_{s-1} } =0$,
$T_{A_1 \ldots A_{s-3}C}{}_{C,}{}^{B_1\ldots B_{s-1} } =0$, and
contains the $o(d-1,2)$ basis elements $T_{A,B}=-T_{B,A}$ such that
\be
[T_{C,D} ,T_{A_1 \ldots A_{s-1},B_1\ldots B_{s-1} }] =
\eta_{DA_1 } T_{C A_2 \ldots A_{s-1}, B_1\ldots B_{s-1} }
+\ldots\,.
\ee

The question is whether there exists  a non-Abelian algebra $h$
with these properties. If yes, the Abelian curvatures $R_1$
(\ref{R1A}) can be understood as resulting from the linearization
of the non-Abelian field curvatures $R$ of $h$ with the $h$ gauge
connection $\tilde{\omega} = \go_0 +\go$, where $\go_0$ is some
fixed flat zero-order connection of the $AdS_d$ subalgebra
$o(d-1,2)\subset h$ and $\go$ is the first-order dynamical part
which describes massless fields of various spins. According to the
discussion of section \ref{Gravity as o(d-1,2) gauge theory}, any
$h$ of this class is a candidate for a global HS algebra of the
symmetric vacuum of a HS theory. The existence of such an algebra
was indicated by the results of \cite{KVZ} where conserved
currents (and therefore charges) in the free massless scalar field
theory were shown to be described by various traceless two-row
rectangular Young tableaux of the conformal algebra. A formal
definition of $h$ as a conformal HS algebra of symmetries of a
scalar field theory in $d-1$ dimension was given by Eastwood in
\cite{East}. In this paper we use a slightly different definition
of $h$ which is more suitable for the analysis of the HS
interactions.

Consider oscillators $Y_i^A$ with $i=1,2$ satisfying the
commutation relations
\be
\label{defr}
[Y_i^A , Y_j^B ]_* = \gvep_{ij}\eta^{AB}\,,\qquad \gvep_{ij}= -
\gvep_{ji}\,,\quad \gvep_{12}=1\,,
\ee
where  $\eta^{AB}$ is the invariant metric of
$o(d-1,2)$. (These oscillators can be interpreted
as conjugated coordinates and momenta $Y^A_1 = P^A$,
$Y^B_2 = Y^B$.)
$\eta_{AB}$ and $\gvep^{ij}$ are used to raise and
lower indices in the usual manner
$
A^A = \eta^{AB} A_B
$,
$
a^i =\gvep^{ij}a_j
$,
$
a_i =a^j \gvep_{ji}\,.
$

We use the Weyl (Moyal) star product
\be
\label{wstar}
(f*g)(Y) =\f{1}{\pi^{2(d+1)}}
\int dS dT f(Y+S) g(Y+T)\exp -2 S^A_i T_A^i\,.
\ee
$[f ,g ]_* = f*g - g*f$, $\{f,g\}_* = f*g + g*f$.
The associative algebra of polynomials
with the $*$ product law generated via
(\ref{defr}) is called Weyl algebra $A_{d+1}$. Its generic
element is
$f(Y) = \sum \phi^{i_1 \ldots i_n}_{A_1 \ldots A_n} Y_{i_1}^{A_1}\ldots
Y_{i_n}^{A_n}\,$
or, equivalently,
\be
\label{exp}
f(Y) = \sum_{m,n} f_{A_1 \ldots A_m\,,B_1 \ldots B_n} Y_{1}^{A_1}\ldots
Y_{1}^{A_m}Y_{2}^{B_1}\ldots
Y_{2}^{B_n}\,
\ee
with the coefficients $f_{A_1 \ldots A_m\,,B_1 \ldots B_n}$
symmetric in the indices $A_i$ and $B_j$.

Various bilinears of oscillators $Y_i^A$ form the
Lie algebra $sp(2(d+1))$ with respect to the star commutator.
It contains the subalgebra $o(d-1,2)\oplus sp(2)$ spanned by the
mutually commuting generators
\be
\label{t}
T^{A,B} = -T^{B,A} =\half  Y^{iA} Y^B_i\,,\qquad
t_{ij} =t_{ji} = Y^A_i Y_{jA}  \qquad [T^{A,B}\,,t_{ij} ]_* =0\,.
\ee
Consider the subalgebra
$S\in A_{d+1}$ spanned by the $sp(2)$ singlets $f(Y)$
\be
\label{sp2}
f\in S:\qquad [t_{ij} , f(Y) ]_* =0\,.
\ee
Eq.(\ref{sp2}) is equivalent to
$
\Big(Y^{Ai} \f{\p}{Y^A_j}  + Y^{Aj} \f{\p}{Y^A_i}
\Big) f(Y) =0\,.
$
For the expansion (\ref{exp})  this condition implies that the
coefficients $f_{A_1 \ldots A_m\,,B_1 \ldots  B_n}$ are nonzero
only if $n=m$ and that symmetrization over any $m+1$ indices of
$f_{A_1 \ldots A_m\,,B_1 \ldots  B_m}$ gives zero, i.e.
$f_{A_1 \ldots A_m\,,B_1 \ldots  B_m}$
has the symmetry properties of a two-row rectangular Young tableau.
The algebra $S$ is not simple.
It contains the two-sided ideal $I$ spanned by the elements
of the form
$
g=t_{ij}*g^{ij}
$,
where $g^{ij}$ transforms as a symmetric tensor with respect to
$sp(2)$, i.e.,
$
[t_{ij}\,,g^{kl}]_* = \delta_i^k g_j{}^l +\delta_j^k g_i{}^l
+\delta_i^l g_j{}^k +\delta_j^l g_i{}^k
$.
(Note that $t_{ij}*g^{ij} =
g^{ij} *t_{ij}$.)
Actually, from (\ref{sp2}) it follows that
$f*g,\, g*f \in I$\, $\forall f\in S$, $g\in I$.
  Due to the definition (\ref{t}) of $t_{ij}$,
the ideal $I$ contains all traces
of the two-row Young tableaux. As a result, the algebra
$S/I$ has only traceless two-row tableaux in the expansion (\ref{exp}).

Now consider the Lie algebra with the commutator in  $S/I$ as the
product law. Its real form corresponding to a unitary HS theory in
$AdS_d$ is called
 $hu(1/sp(2)[d-1,2])$ \cite{d}.   Note that, by construction,
the $AdS_d$ algebra $o(d-1,2)$ with the generators $T^{A,B}$
is the subalgebra of $hu(1/sp(2)[n,m])$.

The gauge fields of  $hu(1/sp(2)[d-1,2])$  are
\be
\label{gaug}
\go(Y|x) = \sum_{l=0}^\infty \go_{A_1 \ldots A_l\,,B_1 \ldots B_l}(x)
 Y_{1}^{A_1}\ldots Y_{1}^{A_l}Y_{2}^{B_1}\ldots Y_{2}^{B_l}
\ee
with the component gauge fields $\go_{A_1 \ldots A_l\,,B_1 \ldots
B_l}(x)$ taking values in all traceless two-row rectangular Young tableaux
of $o(d-1,2)$. Note that, because $dt_{ij} =0$,
the $sp(2)$ invariance condition, which imposes
 the Young symmetry properties, can be
written in the covariant form
\be
\label{covhowe}
D(t_{ij}) \equiv dt_{ij} + [\go ,t_{ij} ]_* =0\,.
\ee

The HS curvatures and gauge transformations have the standard
Yang-Mills form
\be
\label{HSC}
R(Y|x) = d\go (Y|x) +\go(Y|x)\wedge *\go(Y|x)\,,
\ee
\be
\delta \go (Y|x) = D \gvep (Y|x) \,,\qquad
 D \gvep (Y|x) = d\gvep (Y|x) +[\go(Y|x), \gvep (Y|x)]_* \,.
\ee

Different spins correspond to irreducible representations of
$o(d-1,2)$ spanned by homogeneous polynomials
\be
\label{scale}
\go(\mu Y |x ) = \mu^{2(s-1)} \go (Y|x)\,.
\ee
(Note that one unit of spin is carried by the 1-form index). In
particular, spin 1 is described by a 1-form $\go(x) = dx^\un \go_\un (x)$.
The algebra  $hu(1/sp(2)[d-1,2])$ is infinite dimensional.
It contains $o(d-1,2)\oplus u(1)$ as the maximal finite dimensional
subalgebra with the generators $T^{AB}$ (\ref{t}) for
$o(d-1,2)$ and  constants for $u(1)$. The corresponding gauge fields
carry spins 2 and 1, respectively.
Taking two HS symmetry parameters $\gvep_{s_1}$ and $\gvep_{s_2}$,
being polynomials of degrees $s_1 -1$ and $s_2-1$, respectively,
one obtains
\be
\label{ee}
[\gvep_{s_1} \,,\gvep_{s_2} ]_*= \sum_{t=|s_1 - s_2|+1}^{s_1 + s_2 -2}
\gvep_t \,.
\ee
Thus, once a spin $s>2$ gauge field
appears, the HS symmetry algebra requires an infinite tower of
HS gauge fields to be present. The barrier $s\leq 2$  separates
theories with infinite dimensional gauge symmetries from those with
usual lower spin symmetries. It is tempting to speculate that the latter
result from the spontaneous breaking of infinite dimensional HS
symmetries down to usual lower spin symmetries. In that case, the HS gauge
fields should acquire masses as a result of this spontaneous HS symmetry
breaking.

{}The formula (\ref{ee}) manifests the quantum-mechanical nonlocality
of the oscillator algebra (\ref{defr}) (equivalently, the star product
 (\ref{wstar})). Because bilinear terms in the HS curvatures
(\ref{HSC}) describe interactions, one concludes that lower spins
form sources for higher spins and vice versa. A less obvious fact,
which follows from the HS field equations, is that the nonlocal character
of the star product algebra results in the appearance of higher
space-time derivatives in the HS interactions. Thus the
star-product origin of the HS algebra links together such seemingly
different properties of the HS theories as the relevance of the
{$AdS$} background, necessity of introducing infinitely many spins
and space-time non-locality of the HS interactions. Note
that these properties make the HS theories reminiscent
of the superstring theory with the analogy between the cosmological constant
and $\alpha^\prime$.

To introduce inner symmetries, one considers following \cite{KV1}
matrix-valued gauge fields $ \go(Y|x) \longrightarrow\go_\nu{}^\mu
(Y|x) $, $\mu,\nu\ldots = 1,\ldots , p$, imposing the reality
condition \be [\go_\nu{}^\mu (Y|x)]^\dagger = -\go_\nu{}^\mu (Y|x)
\,, \ee where the involution $\dagger$ combines matrix hermitian
conjugation with the involution of the star product algebra
$(Y^A_j)^\dagger = i Y^A_j $. The resulting real Lie algebra  is
called $hu(p |sp(2)|[d-1,2] )$. Its gauge fields  describe the set
of massless fields of all spins $s\geq 1$ which
 take values in the adjoint
representation of $u(p)$. In particular, spin 1 gauge fields are $u(p)$
Yang-Mills fields.

Combining the
antiautomorphism of the star product algebra
$\rho (f(Y)) = f (iY)$ with some antiautomorphism of the matrix
algebra generated by a nondegenerate form $\rho_{\ga\gb}$
one can impose the conditions \cite{KV1}
\be
\go_\ga{}^\gb  (Y|x)=
- \rho^{\gb\gga}\rho_{\delta\ga} \go_\gga{}^\delta (iY|x)\,,
\ee
which
truncate the original system to the one with the Yang-Mills
gauge group $USp(p)$ or $O(p)$ depending on whether the form
$\rho_{\ga\gb}$ is antisymmetric or symmetric, respectively.
The corresponding global HS symmetry algebras are called
$husp(p|sp(2)[d-1,2])$ and $ho(p|sp(2)[d-1,2])$, respectively.
In this case  all fields of odd spins take values in the
adjoint representation of the Yang-Mills group while fields
of even spins take values in the opposite symmetry second
rank tensor representation (i.e., symmetric for $O(p)$ and
antisymmetric for $USp(p)$) which contains a singlet. The graviton
is always the color singlet. For general $p$, color spin 2 particles
also appear however. Note that this
does not contradict to the no-go results of
\cite{CW} because the theory under consideration
does not allow a flat limit with  unbroken
HS  and color spin 2 symmetries. The
minimal HS theory is  based on the algebra $ho(1|sp(2)[n,m])$.
It describes  even spin particles, each in one copy. (Odd spins
do not appear because the adjoint representation of $o(1)$ is trivial.)

\section{Unfolded higher spin dynamics}
\label{Unfolded higher spin dynamics}

An efficient approach to HS dynamics consists of reformulation of
linear and non-linear field equations in the form of some
generalized covariant constancy conditions first at the linear and
then at the nonlinear level. This ``unfolded formulation"
originally introduced for the description of $4d$ HS dynamics in
\cite{Ann} allows one to control simultaneously formal consistency
of field equations, gauge symmetries and the invariance under
diffeomorphisms. The unfolded formulation treats uniformly higher
derivatives of the dynamical fields and is just appropriate for
the analysis of HS dynamics because HS symmetries mix higher
derivatives of the dynamical fields.  The same time, the unfolded
formulation is a universal tool applicable to any dynamical system
although it may be looking  unusual from the perspective of
standard field theory because it operates in terms of infinite
dimensional modules which describe all degrees of freedom of the
system in question. To illustrate the idea let us  recall the
unfolded formulation of the Einstein gravity, using the
compensator formalism.

As explained in section \ref{Gravity as o(d-1,2) gauge theory},
Riemann tensor and torsion tensor are components of the $o(d-1,2)$
field strength (\ref{R})
$R^{AB} = dx^\un \wedge dx^\um R_{\un\um}{}^{AB}$.
The components of the Riemann tensor, that can be nonzero when
Einstein equations and zero-torsion constraints are satisfied,
belong to the Weyl tensor, i.e. Einstein
equations with the cosmological term
can be rewritten as
\be
\label{oms2}
R^{A,B}\Big |_{o.m.s.} = E_C \wedge E_D C^{AC,BD}\,,
\ee
where $C^{AC,BD}$ is treated as an independent
tensor field variable that has the symmetry properties
of the window Young tableau
\quad
\begin{picture}(13,12)(0,0)
{\linethickness{.250mm}
\put(00,10){\line(1,0){10}}
\put(00,05){\line(1,0){10}}
\put(00,00){\line(1,0){10}}
\put(00,00){\line(0,1){10}}
\put(05,00){\line(0,1){10}}
\put(10,00){\line(0,1){10}}
}
\end{picture}
and describes the Weyl tensor. For our purpose it is
convenient to use the symmetric basis with
$C^{AC,BD}=C^{CA,BD}=C^{AC,DB}$. In addition, $C^{AC,BD}$ has the
following properties
\bee
\label{0t}
\mbox{ zero-torsion constraint}:&{}&\qquad
V_A C^{AB,CD} =0\,,\\
\label{EE}
\mbox{ Einstein equations}:&{}&\qquad
C_B{}^{B,CD} =0\,,\\
\label{BI}
\mbox{Bianchi identities for (\ref{0t})}:&{}&\qquad
C^{\{A_1 A_2,A_3\} D} =0:     \qquad
\begin{picture}(13,12)(0,0)
{\linethickness{.250mm}
\put(00,10){\line(1,0){10}}
\put(00,05){\line(1,0){10}}
\put(00,00){\line(1,0){10}}
\put(00,00){\line(0,1){10}}
\put(05,00){\line(0,1){10}}
\put(10,00){\line(0,1){10}}
}
\end{picture}.
\eee

Field equations for free totally  symmetric integer spin $s\geq 2$
massless  HS fields in $AdS_d$  \cite{LV} can analogously be
rewritten in the form \cite{LV} \be \label{ccomt} R_1^{A_1 \ldots
A_{s-1}, B_1\ldots B_{s-1} }\Big |_{o.m.s.}= E_{0\,A_s} \wedge
E_{0\,B_s} C^{A_1 \ldots A_{s}, B_1\ldots B_{s} }\,.
\ee
The
generalized Weyl tensors $C^{A_1 \ldots A_{s}, B_1\ldots B_{s} }$
are described by the traceless $V^A$--transversal two-row
rectangular Young tableaux of length $s$, i.e., $ V_{A_1} C^{A_1
\ldots A_{s}, B_1\ldots B_{s} }=0, $ $ \eta_{A_1 A_2} C^{A_1
\ldots A_{s}, B_1\ldots B_{s} }=0 $ and $
 C^{\{A_1 \ldots A_{s},A_{s+1}\} B_2\ldots B_{s} }=0 .
$
The equation (\ref{ccomt})
referred to as First On-Mass-Shell Theorem is a consequence of
the massless field equations along with the constraints on auxiliary
and extra fields imposed by requiring appropriate components of
HS curvature to vanish \cite{LV}. Let us note that although the extra
fields $\go_\un{}^{a_1 \ldots a_{s-1}, b_1\ldots b_t }$ with $t\geq 2$
do not contribute to the free action, they do contribute at the
interaction level. To make such interactions meaningful, one
has to express the extra fields in terms of the dynamical
ones modulo pure gauge degrees of freedom. This is achieved by imposing
appropriate constraints \cite{LV} contained in
(\ref{ccomt}) like the torsion constraint in gravity is
contained in (\ref{oms2}).

The Bianchi identities $D_0 (R_1 )=0$ along with  the equation
(\ref{ccomt}) impose some
differential restrictions on the generalized Weyl tensor
$C^{A_1 \ldots A_{s}, B_1\ldots B_{s} }$.
The trick is to denote the components of the first derivatives of
$C^{A_1 \ldots A_{s}, B_1\ldots B_{s} }$, that are allowed to
 be non-zero by the Bianchi identities,
by a new tensor field $C_1$, writing symbolically $D^\L_0 C =
E_0\wedge C_1$, where $D^\L_0$ is the Lorentz derivative and $E_0$
is the $AdS$ frame 1-form. The Bianchi identities for this
equation impose differential conditions on $C_1$ to be written as
$D^\L_0 C_1 = E_0\wedge C_2$, etc. It turns out that the full set
of the 0-forms $C_i$ consists of all two-row traceless $V^A
-$transversal Young tableaux $C^{A_1 \ldots A_{u}, B_1\ldots B_{s}
}$ with the second row of length $s$, i.e., 
\be 
\label{condc}
V_{A_1}C^{A_1 \ldots A_u, B_1\ldots B_{s} }      =0\,,\quad
\eta_{A_1 A_2} C^{A_1 \ldots A_u, B_1\ldots B_{s} }=0 \,,\qquad
C^{\{A_1 \ldots A_{u},A_{u+1}\} B_2\ldots B_{s} } =0\,. 
\ee 
The fields $C^{A_1 \ldots A_{u}, B_1\ldots B_{s} }$ form a basis of
the space of  on-mass-shell nontrivial derivatives of order $u-s$
of the spin $s$ generalized Weyl tensor  $C^{A_1 \ldots A_{s},
B_1\ldots B_{s} }$. The full set of the compatibility conditions
of the equations (\ref{ccomt}) can be written in the form of the
covariant constancy condition \cite{d} \be \label{cmt2}
\tilde{D}_0 C^{A_1 \ldots A_{u}, B_1\ldots B_{s} } =0\qquad u\geq
s\,, \ee where $\tilde{D}_0 $ is the $o(d-1,2)$ covariant
derivative in the so  called twisted adjoint representation.

To define the twisted adjoint representation it is useful to
observe that the set of fields $C^{A_1 \ldots A_{u}, B_1\ldots
B_{s} }$ satisfying (\ref{condc}) spans the space isomorphic to
the space of star-product functions  $C(Y|x)$ satisfying
(\ref{sp2}) and with the ideal $I$ factored out to impose the
tracelessness condition. Then $\tilde{D}_0 $ is \cite{d} \be
\label{twad} \tilde{D}_0 = D_0^\L - 2\Lambda  E_0^A V^B \Big (
{}^\pe Y_{A}^i {}^\pa Y_{Bi} - \f{1}{4}\gvep^{ji} \f{\p}{ \p
{}^\pe Y^{ A j}\p {}^\pa Y^{Bi }} \Big )\,, \ee where the Lorentz
covariant derivative is $
D_0^\L = d +  \go_0^{\L\, AB} {}^\pe Y_{Ai}\f{\p}{\p {}^\pe Y^B_i}
$ and we use notations \be A^{A}_i = {}^\pa A^{ A}_i + {}^\pe A^{
A}_i\,,\qquad {}^\pa A^{ A}_i =  \f{1}{V^2} V^A V_B A^B_i\,,\quad
{}^\pe A^{ A}_i =  A^A_i - \f{1}{V^2} V^A V_B A^B_i\, \ee for any
vector $A^A$. The twisted adjoint covariant derivative
(\ref{twad}) commutes with the operator $ N^{tw}= {}^\pe
Y^A_i\f{\p}{\p {}^\pe Y^A_i}- {}^\pa Y^A_i\f{\p}{\p {}^\pa
Y^A_i}\,. $ This means that the equation (\ref{cmt2}) decomposes
into independent subsystems for the sets of fields satisfying $
N^{tw} C = 2s C $ with various integers $s\geq 0$ (the
eigenvalues of the operator $N^{tw}$ are non-negative on the space
of $o(d-1,2)$ tensors with the symmetry property of a two-row
rectangular Young tableau because having more than a half of
vector indices aligned along  $V^A$ would imply symmetrization
over more than a half of indices, thus giving zero). Different
Lorentz irreducible components of the field $C(Y|x)$ with some
fixed $s$ just give the set of components $C^{A_1 \ldots A_{u},
B_1\ldots B_{s} }$ satisfying (\ref{condc}) with $u-s$ being a
number of indices of the length $u$ two-row traceless rectangular
tableau, contracted with the compensator $V^A$. Note that any spin
$s$ submodule of the twisted adjoint $o(d-1,2)$ module,  where the
0-forms $C$ take their values, is infinite dimensional. This is
expected because the components $C$ parametrize all gauge
invariant combinations of on-mass-shell non-zero derivatives of
the higher spin fields.

Equations (\ref{ccomt}) and (\ref{cmt2}) were originally derived
\cite{LV,5d}
by the equivalent reformulation (unfolding) of the free equations of
motion of totally symmetric massless fields of all integer spins
$s\geq 2$ in $AdS_d$ supplemented with some constraints which express
an infinite set of auxiliary variables via higher derivatives of the
dynamical fields. However,
the equation of motion of a massless scalar coincide \cite{SHV} with the
$s=0$ sector of  equation (\ref{cmt2}). Analogously,  equation
(\ref{cmt2}) with $s=1$ imposes the Maxwell equations on the spin 1
potential (1-form) $\go$ on the left hand side of (\ref{ccomt}).

Thus, the equations (\ref{ccomt}) and (\ref{cmt2}) give an
equivalent form of bosonic symmetric massless fields of all spins
in $AdS_d$ for any $d$. We call this important fact Central
On-Mass-Shell Theorem. It is of key importance in many respects
and, in particular, for the analysis of interactions. Central
On-Mass-Shell theorem  can be formulated in the following compact
form \be \label{CMS1} R_1 ({}^\pa Y , {}^\pe Y |x) = \half E_0^A
\wedge  E_0^B \f{\p^2}{\p Y^A_i \p Y^B_j} \gvep_{ij} C(0,{}^\pe
Y|x )\,, \ee \be \label{CMS2} \tilde{D}_0 (C) =0\,, \ee where $
R_1 = d\go + \go_0* \go +\go *\go_0\,, $ $ \tilde{D}_0 (C)= d C +
\go_0 *C - C*\tilde{\go}_0 $ and $\go_0 = \go_0^{AB} (x)
T_{AB}(Y)$ where $\go_0^{AB} (x)$  satisfies (\ref{ADS}) to
describe the $(A)dS_d$  background, and tilde denotes the
$V$-reflection automorphism, i.e., $\tilde{\go}_0(Y|x)
=\go_0(\tilde{Y}|x) = \go_0 ({}^\pe Y-{}^\pa Y|x)$. The
$V$-transversal components of the expansion of the 0-forms
$C(0,{}^\pe Y)$ on the r.h.s. of (\ref{CMS1}) in powers of $Y^A_i$
give rise to HS Weyl 0-forms $C^{a_1 \ldots a_{s}, b_1\ldots b_{s}
}$ on the r.h.s. of (\ref{ccomt}).
 The key fact is that, as one can readily see, the equations (\ref{CMS1})
and (\ref{CMS2}) are consistent, i.e., the application of the covariant
derivative to the l.h.s. of (\ref{CMS1}) and (\ref{CMS2})
does not lead to new conditions.

\section{Nonlinear higher spin classical dynamics}
\label{Nonlinear higher spin classical dynamics}

The problem is to find a  nonlinear deformation of
the equations (\ref{CMS1}) and (\ref{CMS2}) in which
the linearized
curvature and covariant derivative are replaced with the full ones
with $\go=\go_0 +\go_1$ where $\go_0$ describes the background
$AdS_d$ space-time and $\go_1$ describes the dynamical HS gauge fields.
According to the general analysis of \cite{Ann},
where the $4d$ case was considered,
any deformation, formulated
using the exterior algebra formalism, that is
consistent with the Bianchi identities,
describes gauge invariant interactions and is invariant under
diffeomorphisms. Here we formulate following \cite{d}
a system of equations which generates all nonlinear corrections to
the equations (\ref{CMS1}) and (\ref{CMS2}). Before going into
technical details let us stress that the solution we have
found gives a deformation which is unique modulo field redefinitions.
This means that all dimensionless coupling constants can be rescaled
away in the classical HS model like the dimensionless Yang-Mills constant
$g^2 =|\Lambda|^{\frac{d-2}{2}} \kappa^2$ in the classical pure Yang-Mills
theory. The only nontrivial ambiguity that remains is to
consider HS theories with different HS algebras corresponding to
different Yang-Mills groups according
to the classification of section \ref{Higher spin  algebras}.

Roughly speaking, the idea is to describe complicated nonlinear
corrections to HS equations as a solution of some
simple differential type equations with respect to additional
variables. To this end we double a number of oscillators
by introducing additional variables $Z_i^A$. The full system
of equations is formulated in terms of
the fields $W(Z,Y|x)$, $B(Z,Y|x)$ and   $S(Z,Y|x)$,
where $B(Z,Y|x)$ is a 0-form while
$
W(Z,Y|x)=dx^\un W_\un (Z,Y|x)
$
and
$
S(Z,Y|x)=dZ^A_i S^i_A (Z,Y|x)
$
are connection 1-forms in space-time and auxiliary $Z^A_i$ directions,
respectively.
The fields $\go$ and $C$ are identified with the ``initial data''
for the evolution in $Z$ variables as follows
$
\go (Y|x) = W(0,Y|x)
$,
$
 C (Y|x) = B(0,Y|x)
$.
The $Z$ - connection $S$ will be determined in terms
of $B$ modulo gauge ambiguity.
The differentials satisfy the standard anticommutation
relations
$
dx^\un dx^\um = - dx^\um dx^\un\,,
$
$
dZ^A_i dZ^B_j = - dZ^B_j dZ^A_i\,,
$
$
dx^\un dZ^B_j = - dZ^B_j dx^\un
$
and commute to all other variables (from now on we discard the wedge
symbol).

The space of functions $f(Z,Y)$ is endowed
with the star product
\be
\label{star}
(f*g)(Z,Y) =\f{1}{\pi^{2(d+1)}}
\int dS dT f(Z+S,Y+S) g(Z-T, Y+T)
\exp -2 S^A_i T_A^i \,,
\ee
which is associative, normalized so that $1*f =f*1= f$ and
gives rise to the  commutation relations
$
[Y_i^A , Y_j^B ]_* = \gvep_{ij}\eta^{AB}
$,
$
[Z_i^A , Z_j^B ]_* = - \gvep_{ij}\eta^{AB}
$,
$
[Y_i^A , Z_j^B ]_* =0
$.
The star product (\ref{star}) describes a
normal-ordered basis in $A_{2(n+m)}$ with respect to creation
and annihilation operators $Z-Y$ and $Y+Z$, respectively.

Important property of the star product (\ref{star}) is that it
admits the inner Klein operator
\be
{\cal K} = \exp {-2z_i y^i } \,,\qquad
y_i = \f{1}{\sqrt{V^2}} V_B Y^B_i\,,\quad
z_i = \f{1}{\sqrt{V^2}} V_B Z^B_i\,,
\ee
which has the  properties
\be
\label{K}
{\cal K} *f = \tilde{f}*  {\cal K}\,,\qquad {\cal K} *{\cal K}=1\,,
\ee
where  $\tilde{f}(Z,Y) = f(\tilde{Z} , \tilde{Y})$ with
$\tilde {A}^A={}^\pe A^A -{}^\pa A^\pe$ for $A^A = Z^A, Y^A\ldots$.

The full nonlinear system of HS equations is \cite{d}
\be
\label{WW}
dW+W*W =0\,,\qquad
dS+W*S+ S*W =0 \,,\qquad
dB+W*B-B*\tilde{W} =0\,,
\ee
\be
\label{SS}
S*S = -\half( dZ^A_i dZ_A^i  + 4\Lambda^{-1} dz_i dz^{i} B*\K
)\,,\qquad
S*B = B*\tilde{S}\,,
\ee
where $\tilde{S}(dZ , Z,Y) = S(\tilde{dZ} , \tilde{Z},\tilde{Y})$
and $dz_i = \f{1}{\sqrt{V^2}} V_B dZ^B_i$.
In terms of  a noncommutative connection
$
\W = d +W +S
$
 the system (\ref{WW}), (\ref{SS})
reads \be \label{ncmh} \W *\W =  -\half( dZ^A_i dZ_A^i  +
4\Lambda^{-1}  dz_i dz^{i} B*\K )\,,\qquad \W * B = B
*\tilde{\W}\,. \ee We see that   $ dz_i dz^{i} B*\K$ is the only
nonzero component of the  noncommutative  curvature. The
$B$-dependent part of the equation (\ref{SS}) is responsible for
interactions. Note that $B$ has dimension $cm^{-2}$ to match the
Central On-Mass-Shell theorem (\ref{CMS1}) upon identification of
$B$ with $C$ in the lowest order, i.e., $\Lambda^{-1}  B$ is
dimensionless. This form of the nonzero part of the
non-commutative curvature manifests that taking the flat limit may
be difficult in the interacting theory unless  a value of
 the cosmological constant is shifted by a
condensate which breaks  the HS gauge symmetries.

The system is formally consistent in the sense that the associativity
relations $\W*(\W*\W)=(\W*\W )*\W$ and $(\W*\W)*B = B* (\W*\W )$,
equivalent to Bianchi identities,
are respected by the equations  (\ref{WW}), (\ref{SS}).
The only nontrivial part of the consistency check is that for the
relationship $(S*S)*S=S*(S*S)$ in the sector of $(dz_i )^3$ due to the
second term on the r.h.s. of (\ref{SS}) since
$B*\K$ commutes to everything except for  $dz_i $ to which it
anticommutes by the second equation in (\ref{SS}). However,
this does not break down the consistency of the system
because $(dz_i)^3\equiv 0$.
As a result, the equations (\ref{WW}), (\ref{SS}) are consistent
as ``differential''
 equations with respect to $x$ and $Z$ variables.
A related statement is that the equations (\ref{WW}), (\ref{SS})
are invariant under the gauge transformations
\be
\label{gxz}
\delta \W = [\gvep , \W ]_* \,,\qquad \delta B = \gvep * B -
B*\tilde{\gvep}
\ee
with an arbitrary gauge parameter $\gvep (Z,Y|x)$.

To analyze the equations (\ref{WW}), (\ref{SS}) perturbatively
one sets
$
W=W_0 +W_1
$,
$
S= S_0 +S_1
$
and
$
B=B_0 +B_1
$
with the vacuum solution
$
B_0 = 0
$,
$
S_0 = dZ^A_i Z_A^i
$
and
$
W_0 =  \half \go_0^{AB} (x)  Y^i_A Y_{iB}
$,
where $\go_0^{AB} (x)$ satisfies the zero curvature conditions to describe
$(A)dS_d$.
As shown in \cite{d}, the nontrivial part of the system
(\ref{WW}), (\ref{SS}) in the lowest order
has the form (\ref{CMS1}), (\ref{CMS2}). The same time the system
(\ref{WW})-(\ref{SS}) generates all nonlinear corrections
to the unfolded free HS equations (\ref{CMS1}), (\ref{CMS2}).

An intriguing feature of the unfolded formulation of the nonlinear
HS equations is that their nontrivial part (\ref{SS}) has a form
of deformed oscillator algebra equivalent \cite{Aq} to a
two-dimensional fuzzy hyperboloid in the ``auxiliary"
non-commutative space with the coordinates $Z, Y$. The origin of
this fact can be traced back to the condition that the theory must
admit unbroken $sp(2)$ symmetry with some nonlinearly deformed
generators $t_{ij}$ satisfying the covariant constancy condition
(\ref{covhowe}) at the nonlinear level to guarantee that the
interacting HS theory admits interpretation in terms of the same
set of tensor fields as the free theory (see \cite{d} for more
details). According to \cite{Aq}, from (\ref{ncmh}) it follows
that
 the radius of the fuzzy hyperboloid is
$ R_{H^2} (x) \sim (4\Lambda^{-1} B(x) +1)*(4\Lambda^{-1} B(x)
-3). $ Note that it depends via $B$ on a value of the HS curvature
(including the gravitational one) at a given point $x$ of the
space-time base manifold.

\section{Conclusion}
\label{Conclusion}

The main conclusion is that there exists a class of consistent
nonlinear HS gauge theories in anti-de Sitter space of any
dimension. These theories describe totally symmetric bosonic fields
of all integer spins and are fixed uniquely by the HS gauge
symmetry principle modulo the choice of the spin 1 Yang-Mills group
which can be $U(n)$, $O(n)$ or $Sp(2n)$. Global HS symmetries
of the most symmetric vacua of HS gauge theories are certain
star product algebras which exhibit
usual quantum-mechanical  nonlocality in the
auxiliary noncommutative spaces.
The field equations in the HS gauge theory
map this nonlocality to
space-time nonlocality (i.e., higher derivatives)
at the interaction level.
The same time the HS gauge theories remain local
at the linearized level because the space-time symmetries
are realized in terms  of bilinears of auxiliary oscillators.
Let us note that analogously to the non-commutative geometry
framework of string theory,  HS theory is
based on the associative algebras  (i.e., star product and its
matrix extension). For this reason, in particular, there is no
HS model with the spin 1 gauge group $SU(n)$.

Apart from the purely bosonic HS theories described in this
report, a class of supersymmetric HS models is known to exist in
some particular dimensions which admit  twistor realization of the
HS algebras  in terms of spinor oscillators \cite{d4,sup}. (For
more details on the supersymmetric HS gauge theories we refer the
reader to the reviews \cite{gol,SSO}). An interesting feature of
the HS gauge theories formulated in terms of auxiliary
noncommutative twistor space is that at least at the free field
level they exhibit higher symmetries being symplectic extensions
of the usual conformal algebras \cite{F,BLS,BHS}. The formulation
of HS field equations in the form manifestly invariant  under
higher symplectic symmetries (e.g., $sp(8)$ for the $4d$ HS
systems) can be given \cite{BHS} in the generalized space-time
$\M_M$ with matrix coordinates
$X^{\Omega\Lambda}=X^{\Lambda\Omega}$ where $\Omega$ is spinor
index ($\Omega = 1,\ldots 4$ for the $4d$ case so that the
relevant space $\M_4 $ is ten dimensional). The comment we would
like to make is that this formulation gives a simple construction
for nonlocal, nonlinear conserved currents in free $4d$ models by
using the observation of \cite{GV} that conserved currents  to be
integrated over $M$ cycles in $\M_M$, which are bilinear in the
fundamental HS fields $C(X)$ in $\M_M$,
 are themselves equivalent to the fundamental fields in $\M_{2M}$. This allows
building conserved currents $J_2\sim (J_1)^2\sim C^4$ to be integrated
over $2M$ cycles in $\M_M$, etc. {}From the point of view of
the original space-time, these conserved currents look nonlocal,
containing double integration
\be
Q_1 =\int_{\E^M}J_1 (C)\,,\qquad
Q_2 =\int_{\E^M \times \E^M}J_2(C)\,,\ldots\,,
\ee
where $\E^M$ is an analog of the Cauchy surface in $\M_M$ called
local Cauchy bundle in \cite{mar}. Thus,
the formulation of \cite{BHS,GV}
makes it elementary to generate nonlinear nonlocal
conserved currents for massless fields of all spins
by virtue of the extension of Minkowski space-time to
$\M_M$. One of the interesting problems for the future is
to compare the structure of the nonlocal conserved charges
resulting from this construction with the Yangian generators
discussed recently in the context of the $\N=4$ SYM theory in \cite{DNW}.

Finally, let us list some open problems important for elucidating
the structure of HS gauge theories in higher dimensions and their
possible relationship with string theory.
\begin{itemize}
\item
To develop a theory of mixed symmetry (i.e. neither totally symmetric nor
totally antisymmetric) HS gauge fields in $AdS_d$. This problem is not
completely trivial in view of the observation of \cite{BMV} that
not every free
mixed symmetry field in Minkowski space allows a smooth deformation to
$AdS_d$. The origin of this phenomenon is \cite{Met} that
some of the HS gauge symmetries of the flat space mixed symmetry fields
turn out to be broken in the $AdS$ background.
This means that the covariant description of generic massless
fields in $AdS_d$ requires separate investigation compared to the
 flat space \cite{mixed}.
A progress in this direction achieved recently
in \cite{Zin,ASV,almed} indicates that the general problem has a good chance
to be solved soon. The formulation of
\cite{ASV} in terms of HS gauge potentials analogous to that
used to describe symmetric HS fields looks particularly
promising from the perspective of elucidating a structure of
HS algebras underlying HS gauge theories with mixed symmetry fields.
\item
As a nontrivial consistency test,  it is instructive
to check whether the HS algebras admit unitary representations with the
spectra of spins of massless fields equivalent to those predicted by
the consistent nonlinear field equations (\ref{ncmh}).
\item
To extend the formulation of the nonlinear HS
dynamics to the action level.
\item
To extend the $sp(2M)$ covariant formulation of supersymmetric
HS theories to the nonlinear level.
\item
To find solutions of HS
field equations which break down HS gauge symmetries and introduce
a massive parameter $m$ different from the cosmological
constant. This is necessary to define a low energy expansion in
$\left (\f{1}{m} \f{\p}{\p x}\right )^p$. Note that the cosmological
constant $\Lambda$ cannot be used for this purpose
because the dimensionless operator $\Lambda^{-\half} D$ is of order
one as a consequence of (\ref{DDL}), i.e. an expansion in powers of
derivatives in $AdS$ space without any other dimensionful parameter
$m$ has formal meaning.
\end{itemize}

{\bf Acknowledgement} This research was supported in part by grants
INTAS No.03-51-6346, RFBR No.02-02-17067 and LSS No.1578.2003-2.

\end{document}